%% file: main.tex
\documentclass[conference]{IEEEtran}
\IEEEoverridecommandlockouts

\usepackage{fix-cm}
\usepackage[T1]{fontenc}
\usepackage{amsmath}
\usepackage{cite}
\usepackage{graphicx}
\usepackage{placeins}
\usepackage{stfloats}
\usepackage{balance}
\usepackage{xcolor}
\usepackage{url}
\usepackage{tikz}
\usetikzlibrary{arrows.meta,backgrounds,calc,decorations.pathreplacing,
                fit,matrix,patterns,positioning,shapes.geometric}
\usepackage{pgfplots}
\usepackage{pgfplotstable}
\pgfplotsset{compat=1.18}
\usepackage[hidelinks]{hyperref}

\input{figures/styles}

\title{Release Sooner, Expose Less:\\
Offloading MoE Dispatch in NCCL-EP\\
under Dual-Batch Overlap}

\author{
\IEEEauthorblockN{Hsin-Chun Yin, Khaled Hamidouche, Ian Ziemba,
Arjun Kashyap, Pak Markthub,\\
Subhadeep Bhattacharya, Ke Wen, Assaf Ravid,
Manjunath Gorentla Venkata, and Gil Bloch}
\IEEEauthorblockA{NVIDIA\\
\{hsinchuny, khamidouche, iziemba, arjkashyap, pmarkthub,\\
subhadeepb, kwen, aravid, manjunath, gil\}@nvidia.com}
}

\IEEEpubid{\makebox[\columnwidth]{\copyright\ 2026 NVIDIA. All rights reserved.\hfill}
  \hspace{\columnsep}\makebox[\columnwidth]{}}

\begin{document}
\maketitle
\pagestyle{plain}

\input{sections/00_abstract}
\input{sections/01_introduction}
\input{sections/02_background}
\input{sections/05_method}
\input{sections/06_results}
\input{sections/08_discussion}
\input{sections/09_related_work}
\input{sections/10_conclusion}

\bibliographystyle{IEEEtran}
\balance
\bibliography{bibliography/references}

\clearpage
\onecolumn
\input{sections/11_appendix}

\end{document}

%% file: figures/styles.tex
\definecolor{gpufill}{RGB}{248,206,204}
\definecolor{gpudraw}{RGB}{184,84,80}
\definecolor{cpufill}{RGB}{225,213,231}
\definecolor{cpudraw}{RGB}{150,105,160}
\definecolor{nicfill}{RGB}{213,232,212}
\definecolor{nicdraw}{RGB}{130,179,102}
\definecolor{stagefill}{RGB}{245,245,245}
\definecolor{stagedraw}{RGB}{150,150,150}
\definecolor{gdakicolor}{RGB}{90,90,90}
\definecolor{correctcolor}{RGB}{31,119,180}
\definecolor{freeflowcolor}{RGB}{230,126,34}
\definecolor{deepepcolor}{RGB}{148,103,189}

\newlength{\servingmatrixwidth}
\newlength{\servingplotwidth}
\newlength{\servingpanelheight}
\tikzset{
  gpu/.style={draw=gpudraw, fill=gpufill, rounded corners=1pt},
  cpu/.style={draw=cpudraw, fill=cpufill, rounded corners=1pt},
  nic/.style={draw=nicdraw, fill=nicfill, rounded corners=1pt},
  stage/.style={draw=stagedraw, fill=stagefill, rounded corners=1pt},
  lane/.style={draw=black!60, fill=white},
  flow/.style={-{Latex[length=1.8mm]}, line width=0.55pt},
  asyncflow/.style={-{Latex[length=1.8mm]}, line width=0.55pt, dashed},
  note/.style={font=\scriptsize, align=center},
  boxtext/.style={font=\scriptsize, align=center, text width=#1},
  boxtext/.default=2.2cm,
}

\pgfplotsset{
  colormap={rdbu}{rgb255=(5,48,97) rgb255=(33,102,172)
    rgb255=(67,147,195) rgb255=(146,197,222) rgb255=(209,229,240)
    rgb255=(247,247,247) rgb255=(253,219,199) rgb255=(244,165,130)
    rgb255=(214,96,77) rgb255=(178,24,43) rgb255=(103,0,31)},
}

\newcommand{\ncclepdispatchabsolute}[2]{%
\begin{tikzpicture}[baseline=(current bounding box.north)]
\begin{axis}[
  title={#2},
  title style={font=\small,yshift=2mm},
  width=0.39\textwidth, height=0.33\textwidth, scale only axis,
  xmin=-0.5, xmax=8.5, ymin=-0.5, ymax=8.5, y dir=reverse,
  axis line style={draw=none}, tick style={draw=none},
  xtick={0,...,8},
  xticklabels={1,2,4,8,16,32,64,128,256},
  ytick={0,...,8},
  yticklabels={4/32,4/64,4/128,
               8/32,8/64,8/128,
               16/32,16/64,16/128},
  xticklabel style={font=\scriptsize,rotate=35,anchor=east},
  yticklabel style={font=\scriptsize},
  xlabel={tokens/rank}, ylabel={GPUs / SM cap},
  label style={font=\scriptsize},
  clip=false,
  colormap={latencyblue}{rgb255=(247,251,255) rgb255=(198,219,239)
    rgb255=(107,174,214) rgb255=(33,113,181) rgb255=(8,48,107)},
  point meta min=0, point meta max=350,
  visualization depends on={value \thisrow{label} \as \celllabel},
  visualization depends on={value \thisrow{white} \as \cellwhite},
  nodes near coords={\ifnum\cellwhite=1\color{white}\fi\celllabel},
  every node near coord/.append style={font=\fontsize{7.1}{7.6}\selectfont,anchor=center},
]
\addplot[matrix plot*,mesh/cols=9,mesh/rows=9,point meta=explicit,
         draw=black!12,line width=0.15pt]
  table[col sep=comma,x=x,y=y,meta=us] {#1};
\draw[black!28] (axis cs:-0.5,-0.5) rectangle (axis cs:8.5,8.5);
\end{axis}
\end{tikzpicture}}

\newcommand{\ncclepdispatchrelative}[2]{%
\begin{tikzpicture}[baseline=(current bounding box.north)]
\begin{axis}[
  title={#2},
  title style={font=\small,yshift=2mm},
  width=0.39\textwidth, height=0.33\textwidth, scale only axis,
  xmin=-0.5, xmax=8.5, ymin=-0.5, ymax=8.5, y dir=reverse,
  axis line style={draw=none}, tick style={draw=none},
  xtick={0,...,8},
  xticklabels={1,2,4,8,16,32,64,128,256},
  ytick={0,...,8},
  yticklabels={4/32,4/64,4/128,
               8/32,8/64,8/128,
               16/32,16/64,16/128},
  xticklabel style={font=\scriptsize,rotate=35,anchor=east},
  yticklabel style={font=\scriptsize},
  xlabel={tokens/rank}, ylabel={GPUs / SM cap},
  label style={font=\scriptsize},
  clip=false,
  colormap name=rdbu,
  point meta min=-100, point meta max=100,
  visualization depends on={value \thisrow{label} \as \celllabel},
  visualization depends on={value \thisrow{white} \as \cellwhite},
  nodes near coords={\ifnum\cellwhite=1\color{white}\fi\celllabel\%},
  every node near coord/.append style={font=\fontsize{7.1}{7.6}\selectfont,anchor=center},
]
\addplot[matrix plot*,mesh/cols=9,mesh/rows=9,point meta=explicit,
         draw=black!12,line width=0.15pt]
  table[col sep=comma,x=x,y=y,meta=pct] {#1};
\draw[black!28] (axis cs:-0.5,-0.5) rectangle (axis cs:8.5,8.5);
\end{axis}
\end{tikzpicture}}

\newcommand{\servingmatrix}[1]{%
\begin{tikzpicture}
\begin{axis}[
  width=\servingmatrixwidth, height=1.25\servingpanelheight, scale only axis,
  xmin=-0.5, xmax=6.5, ymin=-0.5, ymax=2.5, y dir=reverse,
  axis line style={draw=none}, tick style={draw=none},
  xtick={0,...,6}, xticklabels={1,32,64,128,256,512,1000},
  ytick={0,...,2},
  yticklabels={NCCL-EP baseline (ms),DeepEP baseline,GDA-KI Free Flow},
  xticklabel style={font=\scriptsize},
  yticklabel style={font=\scriptsize},
  xlabel={client concurrency},
  label style={font=\footnotesize},
  colormap name=rdbu,
  point meta min=-50, point meta max=50,
  colorbar,
  colorbar style={
    width=0.20cm,
    ylabel={TPOT change vs.\ NCCL-EP baseline (\%)},
    tick style={draw=none},
    yticklabel style={font=\scriptsize},
    ylabel style={font=\scriptsize},
  },
  visualization depends on={value \thisrow{label} \as \celllabel},
  visualization depends on={value \thisrow{white} \as \cellwhite},
  visualization depends on={value \thisrow{y} \as \cellrow},
  visualization depends on={value \thisrow{mean} \as \cellmean},
  nodes near coords={\ifnum\cellwhite=1\color{white}\fi
    \ifnum\cellrow=0
      \textbf{\pgfmathprintnumber[fixed,precision=2,zerofill]{\cellmean}}%
    \else
      \textbf{\celllabel\%}%
    \fi},
  nodes near coords align={center},
  every node near coord/.append style={font=\scriptsize, anchor=center},
  clip=false,
]
\addplot[matrix plot*, mesh/cols=7, mesh/rows=3, point meta=explicit]
  table[col sep=comma, x=x, y=y, meta=pct] {#1};
\draw[pattern=north east lines,pattern color=black!14,draw=black!22]
  (axis cs:0.5,-0.5) rectangle (axis cs:2.5,2.5);
\end{axis}
\end{tikzpicture}}

\newcommand{\ncclepmatrix}[5]{%
\begin{tikzpicture}
\begin{axis}[
  title={#2},
  title style={font=\small,yshift=2mm},
  width=0.39\textwidth, height=0.33\textwidth, scale only axis,
  xmin=-0.5, xmax=9.5, ymin=-0.5, ymax=8.5, y dir=reverse,
  axis line style={draw=none}, tick style={draw=none},
  xtick={0,...,9},
  xticklabels={0,1k,1.5k,2k,2.5k,3k,3.5k,4k,6k,8k},
  ytick={0,...,8}, yticklabels={1,2,4,8,16,32,64,128,256},
  xticklabel style={font=\scriptsize,rotate=35,anchor=east},
  yticklabel style={font=\scriptsize},
  xlabel={covering GEMM dimension}, ylabel={tokens/rank},
  label style={font=\scriptsize},
  colormap={exposedtime}{
    rgb255=(255,255,229)
    rgb255=(254,227,145)
    rgb255=(254,196,79)
    rgb255=(240,140,55)
    rgb255=(217,95,14)
  },
  point meta min=0, point meta max=200,
  visualization depends on={value \thisrow{us} \as \cellus},
  nodes near coords={\shortstack{%
    \pgfmathprintnumber[fixed,precision=0]{\cellus}\\[-0.5pt]
    {\fontsize{5.7}{6.0}\selectfont$\mu$s}}},
  every node near coord/.append style={
    font=\fontsize{7.1}{7.6}\selectfont,anchor=center,text=black},
  #3
]
\addplot[matrix plot*, mesh/cols=10, mesh/rows=9, point meta=explicit,
         draw=black!12,line width=0.15pt]
  table[col sep=comma, x=x, y=y, meta=us] {#1};
\if\relax\detokenize{#4}\relax\else
  \foreach \x/\y in {#4} {
    \pgfmathsetmacro{\cellleft}{\x-0.48}
    \pgfmathsetmacro{\cellright}{\x+0.48}
    \pgfmathsetmacro{\celltop}{\y-0.48}
    \pgfmathsetmacro{\cellbottom}{\y+0.48}
    \edef\drawcoveragecell{%
      \noexpand\draw[black,line width=0.8pt]
        (axis cs:\cellleft,\celltop) rectangle
        (axis cs:\cellright,\cellbottom);%
    }
    \drawcoveragecell
  }
\fi
\if\relax\detokenize{#5}\relax\else
  \foreach \x/\y in {#5} {
    \pgfmathsetmacro{\cellleft}{\x-0.48}
    \pgfmathsetmacro{\cellright}{\x+0.48}
    \pgfmathsetmacro{\celltop}{\y-0.48}
    \pgfmathsetmacro{\cellbottom}{\y+0.48}
    \edef\drawcensoredcell{%
      \noexpand\draw[black,densely dashed,line width=0.8pt]
        (axis cs:\cellleft,\celltop) rectangle
        (axis cs:\cellright,\cellbottom);%
    }
    \drawcensoredcell
  }
\fi
\end{axis}
\end{tikzpicture}}

\newcommand{\waitexposurematrix}[3]{%
\begin{tikzpicture}
\begin{axis}[
  title={#2}, title style={font=\small,yshift=2mm},
  width=0.39\textwidth, height=0.33\textwidth, scale only axis,
  xmin=-0.5, xmax=10.5, ymin=-0.5, ymax=8.5, y dir=reverse,
  axis line style={draw=none}, tick style={draw=none},
  xtick={0,...,10}, xticklabels={0,4,8,16,32,64,96,128,256,512,1024},
  ytick={0,...,8}, yticklabels={1,2,4,8,16,32,64,128,256},
  xticklabel style={font=\scriptsize,rotate=35,anchor=east},
  yticklabel style={font=\scriptsize},
  xlabel={requested GPU wait ($\mu$s)}, ylabel={tokens/rank},
  label style={font=\scriptsize},
  colormap={waitexposedtime}{
    rgb255=(255,255,229)
    rgb255=(254,227,145)
    rgb255=(254,196,79)
    rgb255=(240,140,55)
    rgb255=(217,95,14)
  }, point meta min=0, point meta max=200,
  visualization depends on={value \thisrow{us} \as \cellus},
  nodes near coords={\shortstack{%
    \pgfmathprintnumber[fixed,precision=0]{\cellus}\\[-0.5pt]
    {\fontsize{5.7}{6.0}\selectfont$\mu$s}}},
  every node near coord/.append style={
    font=\fontsize{7.1}{7.6}\selectfont,anchor=center,text=black}
]
\addplot[matrix plot*, mesh/cols=11, mesh/rows=9, point meta=explicit,
         draw=black!12,line width=0.15pt]
  table[col sep=comma, x=x, y=y, meta=us] {#1};
\foreach \x/\y in {#3} {
  \pgfmathsetmacro{\cellleft}{\x-0.48}
  \pgfmathsetmacro{\cellright}{\x+0.48}
  \pgfmathsetmacro{\celltop}{\y-0.48}
  \pgfmathsetmacro{\cellbottom}{\y+0.48}
  \edef\drawnearminimum{%
    \noexpand\draw[black,line width=0.8pt]
      (axis cs:\cellleft,\celltop) rectangle
      (axis cs:\cellright,\cellbottom);%
  }
  \drawnearminimum
}
\end{axis}
\end{tikzpicture}}

%% file: sections/00_abstract.tex
\begin{abstract}
Dual-batch overlap (DBO) hides mixture-of-experts (MoE) communication behind
another microbatch's computation, but only after dispatch send releases the
GPU.  The time dispatch send holds the GPU therefore stays exposed, while
network progress after release can be hidden.  We evaluate GPUDirect Async -
Kernel Initiated (GDA-KI) Free Flow, an NCCL-EP research configuration that leaves WQE construction on the GPU
but moves send progression to a host thread.  On a GB300 NVL72 system,
GDA-KI Free Flow reduces dispatch-send time relative to default GDA-KI at every
operating point in
the evaluated parameter sweep.  An experiment that inserts a GPU delay
between dispatch send and receive shows that offload increases the in-flight
interval.  Depending on the load, 32--96~$\mu$s of computation or other
independent work between send and receive hides the deferred progress,
leaving less exposed time than default GDA-KI.  In disaggregated vLLM serving
with DBO, GDA-KI Free Flow lowers median TPOT by 15--25\%
relative to NCCL-EP default GDA-KI and by 16--42\% relative to DeepEP v1.
\end{abstract}


%% file: sections/01_introduction.tex
\section{Introduction}
\label{sec:introduction}

Sparse mixture-of-experts (MoE) models route each token to a small subset of
experts distributed across accelerators
\cite{shazeer2017moe,lepikhin2020gshard,fedus2022switch}.  Their dispatch and
combine operations are irregular all-to-all exchanges whose many small,
dynamic transfers make submission latency important during inference.
Under dual-batch overlap (DBO), submission latency is particularly important:
once dispatch send releases the GPU, another microbatch can compute while the
first continues through the host, NIC, and fabric.

NCCL-EP uses NCCL GPU-Initiated Networking (GIN) for its remote transfers.
GIN's default transport backend is GDA-KI.  It removes a CPU launch by having
GPU SMs construct and publish RDMA work-queue entries (WQEs), ring NIC
doorbells, manage credits, and enforce ordering; this work can still block
the sending SM.  This paper seeks to
release the GPU earlier and thereby reduce exposed communication under DBO,
without requiring earlier network completion.

We study GDA-KI Free Flow, a research configuration built on NCCL-EP that
keeps WQE construction on the GPU but moves send progression to a host
thread.  We evaluate this configuration with
NCCL-EP dispatch-send microbenchmarks and with vLLM time per output token
(TPOT) under DBO.

The central mechanism is that \emph{earlier dispatch-send release can reduce
exposed communication even with unchanged or slower network completion,
provided useful work covers the deferred progress}.  In disaggregated vLLM
serving at loads where DBO splits reliably, GDA-KI Free Flow shows 15--25\%
lower median TPOT relative to NCCL-EP
default GDA-KI and 16--42\% lower median TPOT relative to a DeepEP v1
default-GDA-KI baseline.  The dispatch-send microbenchmarks and end-to-end
vLLM serving measurements come from separate experiments; their correlation
is consistent with the proposed mechanism but does not by itself establish
dispatch-send duration as the isolated cause of the TPOT changes.

%% file: sections/02_background.tex
\section{Background and Network-Side Design}
\label{sec:background}

\subsection{MoE Dispatch under Data and Expert Parallelism}

\begin{figure*}[!t]
\centering
\includegraphics[width=0.96\textwidth]{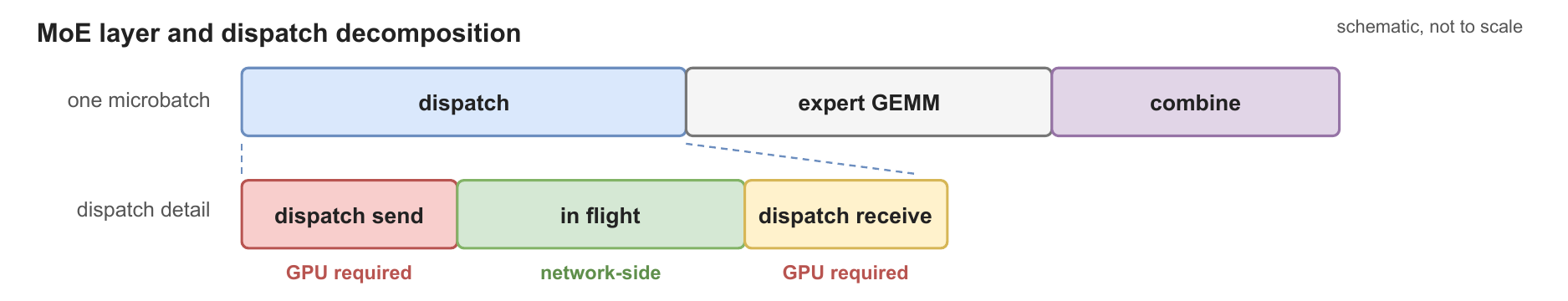}
\caption{An expert-parallel MoE layer.  Each microbatch executes dispatch,
expert GEMM, and combine.  Dispatch consists of GPU-required send and receive
intervals separated by network-side in-flight progress.  Lengths are
schematic.}
\label{fig:moe-workflow}
\end{figure*}

A mixture-of-experts (MoE) layer routes each token to a small subset of its
experts \cite{shazeer2017moe}.  In the deployments considered here, both the
data-parallel (DP) and expert-parallel (EP) dimensions exceed one: DP ranks
own different token batches, while the experts are partitioned across EP
ranks.  A token's selected expert therefore need not be on the GPU that owns
the token.

After routing, the distributed MoE layer has three logical stages.  Dispatch
moves token activations to the ranks that own their selected experts.  Those
experts execute their GEMMs.  Combine then returns the weighted expert
outputs to the source ranks and restores token order.

Figure~\ref{fig:moe-workflow} expands dispatch into three intervals.
\emph{Dispatch send} is the source-GPU work that prepares any required
payloads, incorporates routing information when needed, constructs transfer
requests, and submits the transfers.  Payload preparation may include
packing, but packing is not required by this definition.  \emph{In flight}
is the remaining network-side progress after the source GPU is released.
\emph{Dispatch receive} is the destination-GPU work that waits for arrivals
and makes the received tokens ready for expert computation.  These are
logical ownership boundaries: a transfer may begin before the send kernel
returns.

\subsection{Dual-Batch Overlap and Exposed Communication Time}
\label{sec:exposed-model}

Dual-batch overlap (DBO) splits a batch into two microbatches and schedules
communication from one microbatch beneath computation from the other.
DeepSeek describes this schedule for both prefill and decode as a way to hide
cross-node EP communication \cite{deepseek2025inference}.

\begin{figure*}[!t]
\centering
\includegraphics[width=0.96\textwidth]{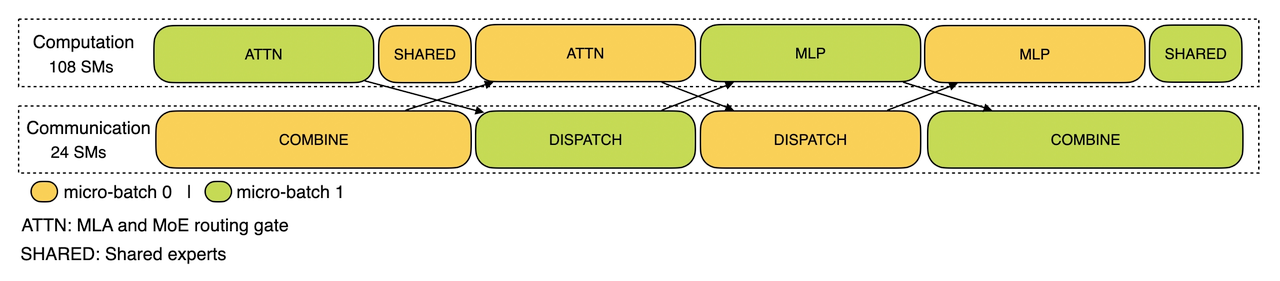}\\[-0.5mm]
\includegraphics[width=0.96\textwidth]{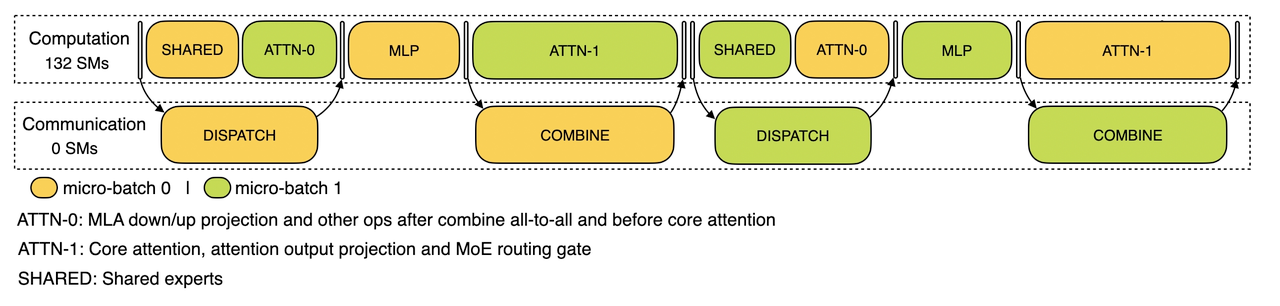}
\caption{DeepSeek's original DBO schedules for prefill (top) and decode
(bottom).  Orange and green identify the two microbatches.  Prefill alternates
their attention, shared-expert, MLP, dispatch, and combine stages.  Decode
splits attention into two parts to form a five-stage pipeline with dispatch
and combine.  In both schedules, communication for one microbatch runs under
computation for the other.  Reproduced from DeepSeek-AI
\cite{deepseek2025inference} (CC0 1.0).}
\label{fig:deepseek-dbo}
\end{figure*}

Figure~\ref{fig:deepseek-dbo} shows the complete schedules that motivate our
model.  In prefill, the longer attention and MLP regions provide covering
windows for dispatch and combine.  Decode has less uniform stage durations,
so DeepSeek splits each microbatch's attention layer into two steps:
\textsc{Attn-0} (the projections before core attention) and \textsc{Attn-1}
(core attention, output projection, and the MoE gate).  Together with
shared-expert and MLP computation, these steps form a five-stage pipeline in
which each stage of one microbatch covers communication of the other.

Our serving experiments exercise the decode schedule.  To identify what a
network backend can improve within that schedule, we next expand one
communication stage into its GPU-required send and receive intervals and its
network-side in-flight interval.

In this paper, \emph{GPU-side submission} denotes a path that keeps transfer
submission and subsequent queue progression on the sending GPU's SMs.
\emph{Network-side submission} denotes a path in which the sending GPU first
publishes the transfer request and, after that publication, the NIC or a host
progress thread performs later progression, allowing the sending GPU to
return earlier.  Both terms describe control ownership after request
preparation; they do not imply different payload or receiver semantics.

\begin{figure*}[!t]
\centering
\includegraphics[width=0.96\textwidth]{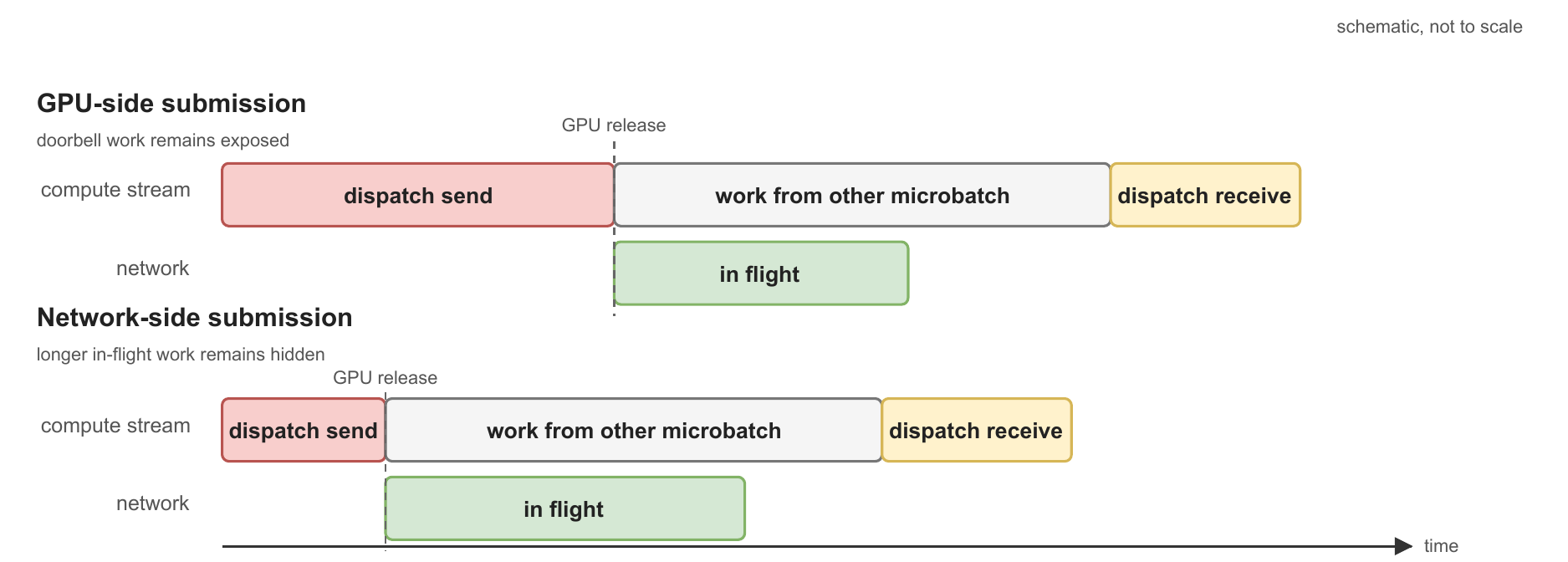}
\caption{Earlier dispatch-send return improves a two-microbatch DBO schedule.
The white block is work from the other microbatch.  Moving later submission
progression away from the sending SM shortens the red dispatch-send interval
and therefore the exposed communication time, starts the covering work
earlier, and keeps the green in-flight interval hidden beneath it.  The
network-side path therefore reaches dispatch receive and completes the MoE
layer earlier even though its illustrated in-flight interval is slightly
longer.  Both rows share one schematic time scale.}
\label{fig:overlap-metric}
\end{figure*}

Figure~\ref{fig:overlap-metric} compares two schedules for the same
two-microbatch MoE layer.  In each row, the red send and yellow receive belong
to one microbatch, while the white block is useful GPU work from the other
microbatch.  The green interval begins exactly when dispatch send releases the
GPU.  With GPU-side submission, the longer send delays both the covering work
and the eventual receive.  Network-side submission returns earlier, so the
other microbatch starts sooner and hides the remaining network progress.  The
lower green interval is deliberately slightly longer than the upper one, yet
it remains shorter than the unchanged white work interval: network completion
need not improve for the MoE layer to finish earlier.

\emph{Exposed communication time} is the portion of communication that remains
on the application critical path after all available computation overlap.
In Fig.~\ref{fig:moe-workflow}, dispatch send and dispatch receive are exposed
because they require the GPU needed by the covering computation.  Once
dispatch send releases that GPU, network-side progress can run beneath the
other microbatch.  In-flight progress contributes to exposed communication
time only when it extends beyond the covering work.

Measure all timestamps from the start of dispatch send.  Let $A$ be the
timestamp at which useful work from the other microbatch is ready, and let
$R_b$ be the timestamp at which backend $b$ releases the sending GPU.  The
work can start only after both events, so its start timestamp is
$S_b=\max(A,R_b)$; the maximum selects the later event.  Thus, $A$, $R_b$,
and $S_b$ are points in time, not durations.

For the baseline and offloaded paths, assume
$R_{\mathrm{off}}\leq R_{\mathrm{base}}$.  Define
$\Delta_R\equiv R_{\mathrm{base}}-R_{\mathrm{off}}$ as the duration by which
offload advances GPU release, and
$\Delta_S\equiv S_{\mathrm{base}}-S_{\mathrm{off}}$ as the duration by which
the covering work starts earlier.  Then
\begin{equation}
0 \leq \Delta_S \leq \Delta_R .
\label{eq:release-bound}
\end{equation}

The advance is full when $A\leq R_{\mathrm{off}}$, partial when
$R_{\mathrm{off}}<A<R_{\mathrm{base}}$, and zero when
$A\geq R_{\mathrm{base}}$.  Thus, earlier GPU release can advance the
covering work by at most the release-time reduction; it need not advance it
at all.  Equation~(\ref{eq:release-bound}) therefore describes only when
covering work can start, not how much network progress it hides.

To model that overlap, let $I_b$ be backend $b$'s in-flight duration from
$R_b$ until the final arrival required by dispatch receive, and let
$T_{\mathrm{cover}}$ be the duration of the other microbatch's covering work.
At $S_b$, the remaining in-flight duration $J_b$ divides into an overlapped
portion $O_b$ and an exposed tail $X_b$:
\begin{equation}
\begin{aligned}
J_b &= [R_b+I_b-S_b]_+, \\
O_b &= \min(T_{\mathrm{cover}},J_b), \qquad
X_b = [J_b-T_{\mathrm{cover}}]_+ ,
\end{aligned}
\label{eq:overlap-tail}
\end{equation}
where $[x]_+=\max(x,0)$.  Dispatch receive can begin at
\begin{equation}
D_b=\max(S_b+T_{\mathrm{cover}},R_b+I_b)
   =S_b+T_{\mathrm{cover}}+X_b .
\label{eq:receive-start}
\end{equation}

When covering work starts at GPU release, $S_b=R_b$ and $J_b=I_b$.  If
$T_{\mathrm{cover}}\geq I_b$, all in-flight progress overlaps and $X_b=0$.
If $T_{\mathrm{cover}}<I_b$, the uncovered duration is
$X_b=I_b-T_{\mathrm{cover}}$.  When $S_b>R_b$, some network progress occurs
before the covering work starts, so $T_{\mathrm{cover}}$ need only cover
$J_b$, not the full $I_b$.  Dispatch send, dispatch receive, and any nonzero
$X_b$ remain exposed.  Reducing dispatch send can therefore shorten the layer
even when offload increases $I_b$, provided the covering work keeps $X_b$
small enough.  Whether DBO supplies enough covering work is an empirical
question, which Sections~\ref{sec:overlap-results}
and~\ref{sec:serving-results} examine.

\subsection{GIN and Its GDA-KI Paths}

\begin{figure*}[!t]
\centering
\includegraphics[width=0.98\textwidth]{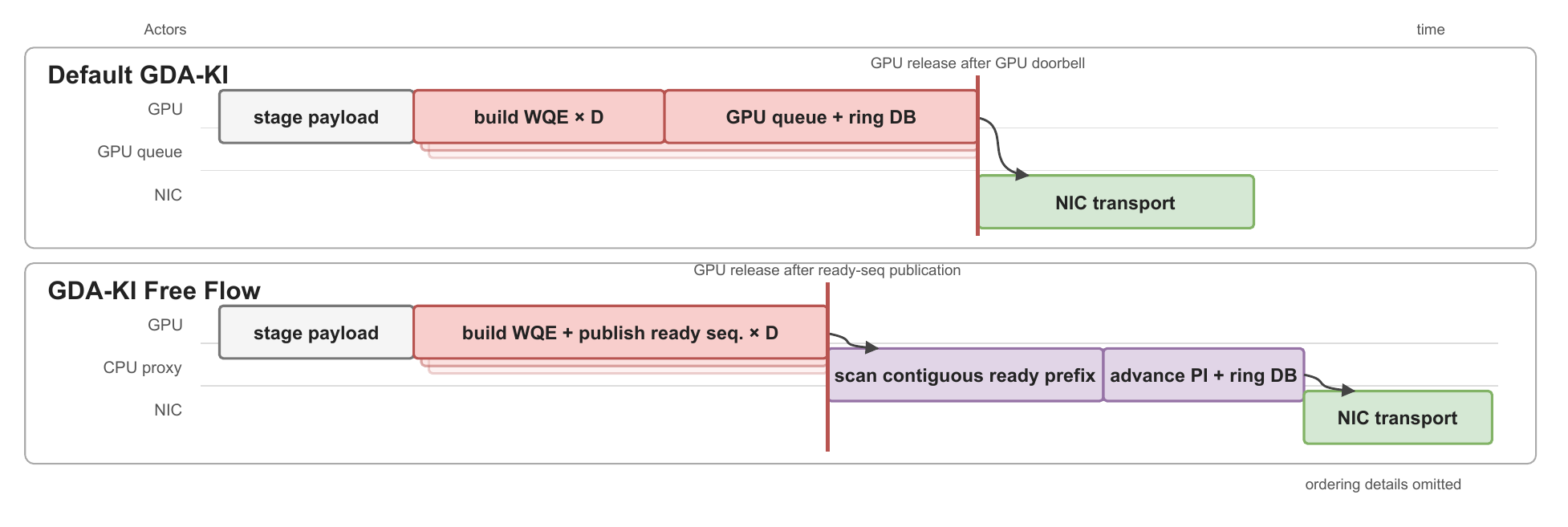}
\caption{Default GDA-KI and GDA-KI Free Flow on a common actor/time layout.
Red spans denote the dispatch-send component of exposed communication time
and are schematic: Free Flow releases the GPU earlier by moving queue
progression and the doorbell to a host thread.  Stacked red boxes denote
repeated per-destination WQE operations, parallelized across GPU threads.
Ordering fences required by both paths are omitted.}
\label{fig:three-paths}
\end{figure*}

GPU-Initiated Networking (GIN) is NCCL's GPU-facing network abstraction for
operations such as device-side memory windows, puts, signals, and
completions.  NCCL-EP uses those operations to implement the remote transfers
in dispatch and combine while selecting among transport backends
\cite{hamidouche2025gin,ncclEp}.

We study GIN's default GDA-KI path and its Free Flow optimization.
Default GDA-KI builds RDMA work-queue entries (WQEs) in GPU memory and performs
the queue progression and NIC doorbell from the GPU.  GDA-KI Free Flow keeps
the GPU-built WQE but moves ready-prefix progression and the doorbell to a
host proxy.  Both paths repeat GPU-side WQE preparation and publication
across destination transfers; the stacked cards in Fig.~\ref{fig:three-paths}
show that these operations are parallelized, with many GPU threads
collaboratively posting the sends.

\subsubsection{Default GDA-KI and GDA-KI Free Flow}

Default GDA-KI places the send queue and WQEs in GPU memory, and GPU code
manages the queue and rings the NIC doorbell directly
\cite{docaGpuNetIO,hamidouche2025gin}.  In the tested NCCL
version (2.31.2), a sending SM reserves queue space, writes the WQE,
advances the send-queue producer state, serializes shared-QP state, and rings
the NIC.  The NIC then fetches the WQE and payload.  Queue management and the
doorbell therefore remain part of dispatch send.

GDA-KI Free Flow changes control ownership without changing the payload path.  The
GPU still reserves a send-queue slot and writes the complete WQE, but then
release-orders that slot's absolute sequence number into its ready-array
entry and returns.  Each ready-array entry marks one individual queue slot
ready; the absolute sequence number prevents a stale marker from a previous
queue lap from validating a reused slot.  A host progress thread derives the
\emph{ready prefix}, defined as the longest contiguous sequence of ready
entries beginning at the current producer frontier, then advances that
frontier through the prefix and rings the NIC.  Thus, the ready prefix is
derived queue state, not an element of the ready array.
This design shortens the GPU-resident send path by shifting queue progression
to the host, and therefore relies on continuously available CPU progress
capacity.

Both paths enforce the two orderings that dispatch correctness requires:
each payload is staged before submission writes a WQE and rings its doorbell;
each payload transfer then precedes
the remote signal that announces it under the transport's ordering semantics.
Violating either would let the receiver read incomplete data.  The
optimization changes only where send-side work progresses, not these
orderings or the receiver-visible dispatch semantics.

\FloatBarrier

%% file: sections/05_method.tex
\section{Method}
\label{sec:method}

We use three sets of experiments.  First, we measure how long each backend
keeps the GPU busy during dispatch send.  Second, we insert a controlled
delay between dispatch send and receive to measure how much time is needed
to hide the remaining network work.  Third, we run vLLM with DBO and measure
end-to-end TPOT.

\begin{table}[!t]
\centering
\caption{Software stack and runtime settings.}
\label{tab:artifact-pins}
\footnotesize
\begin{tabular}{@{}>{\raggedright\arraybackslash}p{0.32\columnwidth}
                    >{\raggedright\arraybackslash}p{0.63\columnwidth}@{}}
\hline
Component & Version or setting \\
\hline
vLLM & 0.30.1rc0-based fork, one-stream DBO \\
NCCL with GIN & 2.31.2-based research forks \\
MoE integration & Research prototype \\
DeepEP baseline & v1.2.1-based LL research fork \\
Model & Qwen3-30B-A3B \texttt{ad44e777}, BF16 \\
Platform & GB300-CX8 NVL72, aarch64, CUDA 13.0 \\
GIN contexts/rank & 4 \\
Serving SM cap & 64 \\
\hline
\end{tabular}
\end{table}

\subsection{NCCL-EP Experiments}
\label{sec:ep-method}

Both experiments use NCCL-EP's low-latency path with BF16 activations,
hidden dimension 2048, 128 experts, four GIN contexts per rank, and NVLink
disabled.  We evaluate default GDA-KI and GDA-KI Free Flow.
Every cell uses equal token counts on all ranks and random top-$k$ expert
selection without replacement.  The routing is computed once per
cell and reused for all timed iterations and rounds, so the sweeps do not
vary routing skew.  Table~\ref{tab:artifact-pins} lists the software
versions and runtime settings.  The dispatch factor sweep keeps Free Flow and its reference on the same
allocation within a recorded shuffled shape block.  The wait sweep uses
one common two-node allocation for both backends in each round, with
backend order shuffled reproducibly and node groups allowed to vary between
rounds.

\paragraph{Experiment 1: dispatch-send factor sweep}
This experiment asks how much each backend reduces dispatch-send time, the
first component of exposed communication time, across MoE operating points.
It crosses GPU count $\{4,8,16\}$,
dispatch SM cap $\{32,64,128\}$, top-$k\in\{4,8\}$, and tokens/rank
$\{1,2,4,8,16,32,64,128,256\}$.
Each iteration launches dispatch send, an 8192-square BF16 GEMM, and dispatch
receive.  The GEMM is only a long covering window before receive; CUPTI
directly measures the dispatch-send kernel reported in
Figure~\ref{fig:ncclep-dispatch-send}.

For each round we take the slowest-rank mean, then average across five
separately executed rounds.  Relative panels first compute the percentage change
against default GDA-KI within the same round and allocation.  Every cell
has five matched rounds; round identities and variation are retained in
the underlying data.

\paragraph{Experiment 2: controlled GPU-wait sweep}
This experiment asks how much time after GPU release reduces the work exposed
at dispatch receive.  We fix EP8 on two four-GPU nodes, top-$k=8$, and dispatch
cap 128, and sweep the same nine token counts.  Requested waits are
$\{0,4,8,16,32,64,96,128,256,512,1024\}$~$\mu$s.  Each sequence launches
separate dispatch-send, GPU-delay, and dispatch-receive kernels on the caller's
compute stream.  PyTorch's built-in delay uses one block and one thread;
cycles are calibrated per GPU, and CUPTI records the actual delay in the
same sequence.  A no-delay sequence and a separate zero-cycle delay-kernel
control record the delay kernel's duration and insertion effects.
Figure~\ref{fig:ncclep-overlap} displays the eleven requested waits, including
the no-delay case; the empty-kernel control is not plotted.

Each point captures 20 sequences in a CUDA graph after three ordinary warmup
iterations.  Three graph warmup replays precede three unprofiled and three
CUDA-only CUPTI-profiled timed replays, retaining 60 paired send/receive
samples per rank.  A captured all-reduce on a dedicated NCCL control group
aligns peer GPUs before external CUDA timing events bracket the 20 sequences;
this alignment is outside the measured window.  The default Gloo host-control
group is not captured.  Receive-count and BF16 round-trip checks, phase counts,
stream placement, and launch geometry are validated for every point.

For each rank we average the same-invocation sum
$T_{\mathrm{send}}+T_{\mathrm{receive}}$, take the largest of the eight rank
means within a round, and report the median of five rounds.  Because we time
the two communication kernels directly, the delay itself is excluded.  Actual delays, signed inter-kernel gaps,
per-rank samples, and complete-sequence timing were also recorded.
All 198 plotted backend/token/wait cells have five accepted rounds.

To check that CUPTI profiling does not distort these timings, the profiled
and unprofiled CUDA-event means of every point must agree within the larger
of 5~$\mu$s or 20\% of the unprofiled mean; all reported points pass.

The GPU delay stands in for covering work without adding a GEMM's compute and
memory traffic.  The exposed-time floor includes send and receive execution
and does not approach zero.  We mark the shortest wait whose measured time is
within 5\% of the lowest value for that backend and token count.

\subsection{vLLM TPOT Experiments}
\label{sec:serving-method}

The native experiments measure GPU release and overlap directly.  The vLLM
experiments test whether real DBO supplies enough work to hide communication
and reduce TPOT.  We plug
the two NCCL-EP decode backends into vLLM and
serve Qwen3-30B-A3B in BF16 on a 16-GPU GB300-CX8 NVL72 allocation: two
four-GPU prefill nodes and two four-GPU decode nodes.  We report median time
per output token (TPOT).  Every point averages three run medians; relative
values use the common three-run default-GDA-KI mean.  The caller compute
stream runs both low-latency send and receive rather than using a separate
communication stream, matching the stream configuration used by DBO.
Both serving roles set the dispatch/combine device-SM cap to 64 with four
GIN contexts per rank, so serving does not run at the cap of 128 used in the native wait
sweep and several native examples.

The workload uses synthetic random prompts with temperature 0 and
\texttt{--ignore-eos}, so every request generates exactly its requested
output length.  Input length is nominal: the prompt generator can yield 127
tokens for a nominal length of 128; saved request records retain actual
lengths.  Each server session runs 64 warmup requests before measurement.
At client concurrency $C$, at most $C$ requests are in flight at once, and
each point measures $4C$ requests (at least 64) with no additional
request-rate limit.

We also include a DeepEP default-GDA-KI serving baseline, using three runs at
each matching concurrency and output length.  The
baseline is a DeepEP v1 low-latency (LL) integration fork based on
DeepEP v1.2.1 and running over NCCL GIN, not the upstream NVSHMEM-based V1
release or DeepEP V2 \cite{deepEP}.  We use V1 LL because
its dispatch returns a receive hook: send and receive are separate calls on
the caller compute stream, the same hook-mode release boundary used by our
NCCL-EP configuration under DBO.  Upstream DeepEP has since replaced V1 with
a unified V2 interface and removed the V1 APIs.  The model, precision, 2P2D topology, DBO
threshold, decode graph mode, and intranode-only NVLink setting match the
NCCL-EP study.  DeepEP uses its default-GDA-KI path on prefill and decode.
The optimized configuration is evaluated within NCCL-EP; DeepEP
contributes only its baseline.  All serving percentages, including the DeepEP row, use the
NCCL-EP default-GDA-KI mean at the same point as their reference.

\paragraph{Experiment 1: concurrency sweep}
This experiment fixes nominal input and output length at 128 tokens and sweeps client
concurrency over $\{1,32,64,128,256,512,1000\}$.  It combines four serving
techniques:

\emph{1) Disaggregated serving.}
We use a 2P2D deployment so prefill and decode execute on separate nodes.
This keeps prefill expert-parallel execution and its EP traffic off the
decode nodes; NIXL KV-cache transfer from prefill to decode still uses the
decode-side network rails and can overlap decode EP traffic.  Every serving
configuration uses default GDA-KI on prefill; the evaluated backend runs on
decode.

\emph{2) CUDA-graph replay.}
Decode uses the V1 model runner and \texttt{FULL\_DECODE\_ONLY} capture.
Replay removes launch overhead and makes the communication interval a larger
fraction of TPOT.  With DBO enabled, the runner captures one CUDA graph per
padded batch shape: a DBO-split graph if the shape reaches the DBO
threshold, and an unsplit graph otherwise.  A step at a split-captured shape
that does not split at run time has no matching graph and falls back to
eager execution; later split steps still replay the captured graph.

This single-graph behavior is intentional: eager execution makes cells that
mix split and unsplit steps visible in TPOT.  Capturing both split and
unsplit graphs would hide that penalty without identifying how much each
path contributes.

\emph{3) RDMA for inter-node traffic.}
NVLink remains enabled within each node, but cross-node NVLink is disabled.
Local traffic therefore uses NVLink while every inter-node expert transfer
uses the RDMA path under test.

\emph{4) Dual-batch overlap.}
DBO uses threshold 2 and two microbatches.  Client concurrency is
cluster-wide, whereas the split decision is made per step from the decode
tokens scheduled on each of the eight data-parallel (DP) decode ranks, before
MoE expert dispatch.  A step splits only if every rank's scheduled count
reaches the threshold and padding all ranks to the largest count would not
leave the final microbatch of the least-loaded rank empty.

We set the threshold to 2 deliberately.  It is the smallest value at which a
step can split into two non-empty microbatches, so DBO engages across the
whole concurrency sweep and its low-load boundary becomes visible.  The sweep
therefore also captures near-threshold steps, such as two scheduled tokens on
a rank split into two one-token microbatches on the same caller compute
stream.  These splits leave almost no compute to cover communication and are
not an economical operating point; a production deployment would use a
higher threshold and would not run them.

Concurrency 1 is an important control: it never activates DBO, so every
step replays its unsplit CUDA graph.  At concurrency 32--64, variation in
request placement and scheduled decode batch sizes can change the all-rank
DBO vote or leave the final microbatch empty after common padding, selecting
the non-DBO path on some steps, while another step at the same padded shape
selects DBO.  Those non-DBO steps run eagerly, as described above, and
\textbf{we exclude these mixed cells from the transport comparison}.  At concurrency
128--1000, each decode rank schedules well above the threshold, and run logs
show both microbatches executing on every decode worker.

\paragraph{Experiment 2: output-length sweep}
This experiment holds concurrency at 256 and nominal input length at 128, then sweeps
output length over $\{128,256,512,1024,2048\}$ with the same two decode
backends.  Each backend has three runs, output lengths increase within each
server session, and every point measures 1,024 requests.
Figure~\ref{fig:serving-output-length} reports the mean of
the three run-level median TPOTs, separately from the concurrency sweep.

\paragraph{Correctness checks}
Because the optimization changes transfer and signal progression, timing is
meaningful only if ordering and outputs remain correct.  Every native cell
checks receive counts and a combine reference.  For
serving, we require zero failed requests and verify the outputs of every
run against DeepEP.  Nine DeepEP runs of the concurrency-256, 128/128 point,
whose 1,024 prompts are identical in every run, form the reference.  Because
batched inference is not bitwise deterministic even at temperature 0, we
compare the first two characters of each generated text, on the 949 prompts
where all nine reference runs agree.  A run passes if at most 3\% of these
prompts differ from the reference.  To confirm that this check tracks task
accuracy, we also evaluated separate runs with GSM8K (1,319 questions,
5-shot, greedy), including a configuration with a known output-corruption
bug: the dma-buf address wrap that the KV-cache alignment guard prevents.

%% file: sections/06_results.tex
\section{Results}
\label{sec:results}

The results answer three questions.  First, does GDA-KI Free Flow
release the GPU sooner than default GDA-KI?  Second, how much independent
work is needed to hide the network progress that remains after release?
Third, when DBO provides that work in vLLM, do the complete configurations
reduce TPOT?  We answer these questions in order with the dispatch-send,
controlled-wait, and serving experiments.

\subsection{NCCL-EP Dispatch-Send Latency}
\label{sec:dispatch-results}

\begin{figure*}[!t]
\renewcommand{\thefigure}{5a}
\centering
\input{figures/ncclep_dispatch_send}

\caption{Five-round NCCL-EP dispatch-send sweep at top-$k=8$.
Columns vary tokens/rank; rows cross GPU count with dispatch CTA/SM cap.
Panel (a) shows default-GDA-KI dispatch-send time in $\mu$s.  Panel (b)
shows mean paired percentage changes for GDA-KI Free Flow,
with a \% sign in every relative cell.  All 162 plotted cells have five matched
rounds.  Relative colors saturate at $\pm100\%$; printed values retain
larger changes.  NVLink is disabled.  The matching top-$k=4$ grid is in
Appendix~\ref{app:dispatch-topk4}, Figure~\ref{fig:ncclep-dispatch-topk4}.}
\label{fig:ncclep-dispatch-send}
\end{figure*}
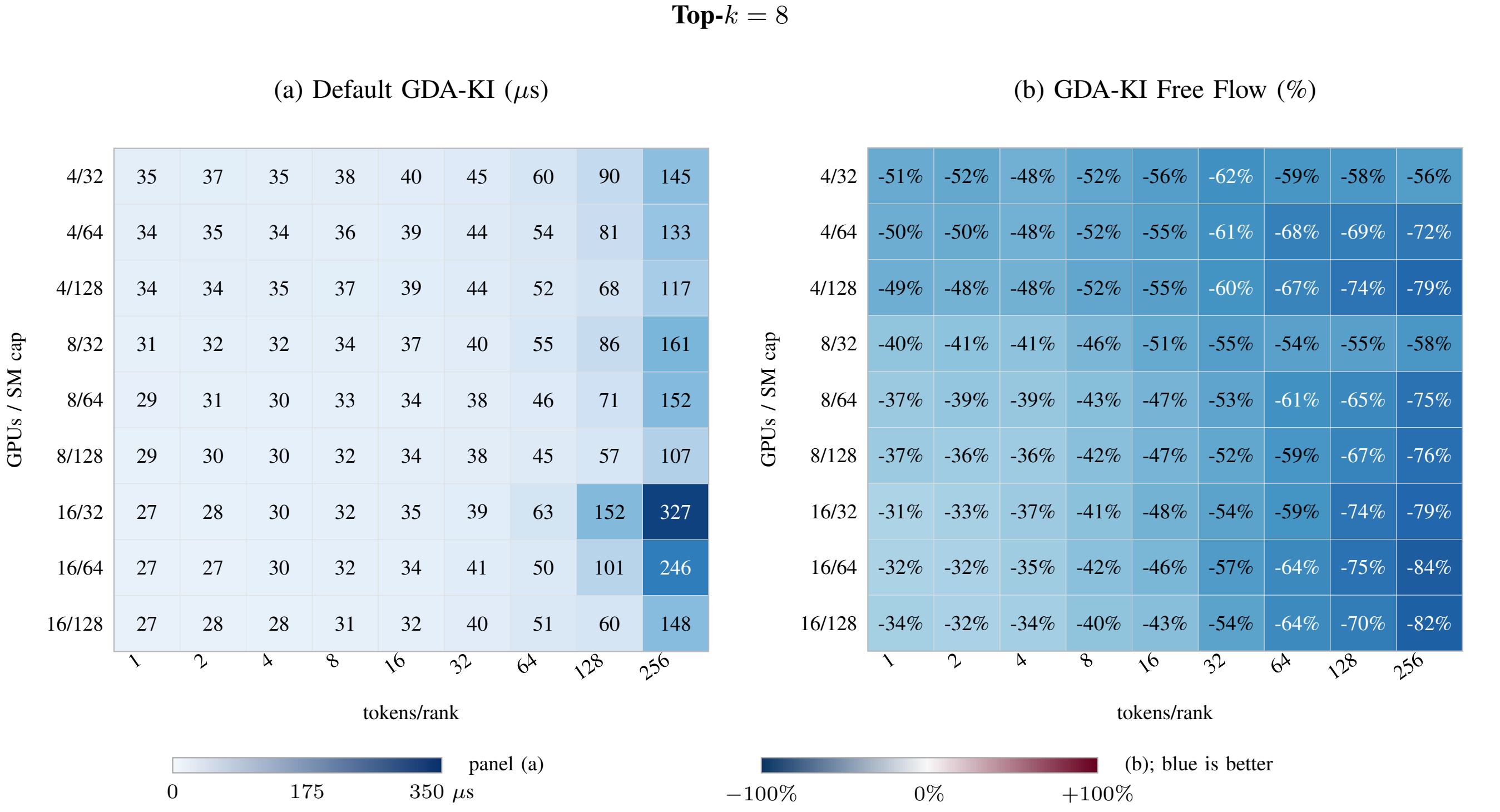

Figures~\ref{fig:ncclep-dispatch-send} and~\ref{fig:ncclep-dispatch-topk4}
report the complete factorial: all 324 backend--position cells across 4, 8, and 16 GPUs have
five matched rounds, 1,620 backend--round measurements in total, and every
cell passes its receive-count and combine-reference checks.

The default-GDA-KI panel shows the mean of the per-round slowest-rank mean
kernel time.  For GDA-KI Free Flow, we first compute its percentage change
from default GDA-KI within the same round and allocation, then average those
paired percentages.  Every comparison now has its same-round baseline,
with the same five rounds used for absolute and relative values.  Negative percentages mean earlier
GPU release.  Ranges and sample counts describe observed rounds rather than
confidence intervals.

\begin{figure*}[!t]
\centering
\input{figures/ncclep_full}
\caption{Five-round NCCL-EP GPU-wait sweep at EP8, top-$k=8$,
dispatch cap 128, and NVLink off.  The two panels show tokens/rank
against requested GPU wait.  Each cell is the median across five rounds of
the slowest-rank mean of directly paired
$T_{\mathrm{send}}+T_{\mathrm{receive}}$, in $\mu$s, excluding the delay
kernel.  Every cell includes its unit.  Zero means no delay kernel; the
separate empty-kernel control is not plotted.  Requested waits are
calibrated per GPU and actual in-sequence delays are recorded.  These are
communication-kernel durations, not differences of standalone timings.
A solid outline marks the first requested wait within 5\% of that row's
lowest observed median; it is an empirical near-floor marker.}
\label{fig:ncclep-overlap}
\end{figure*}
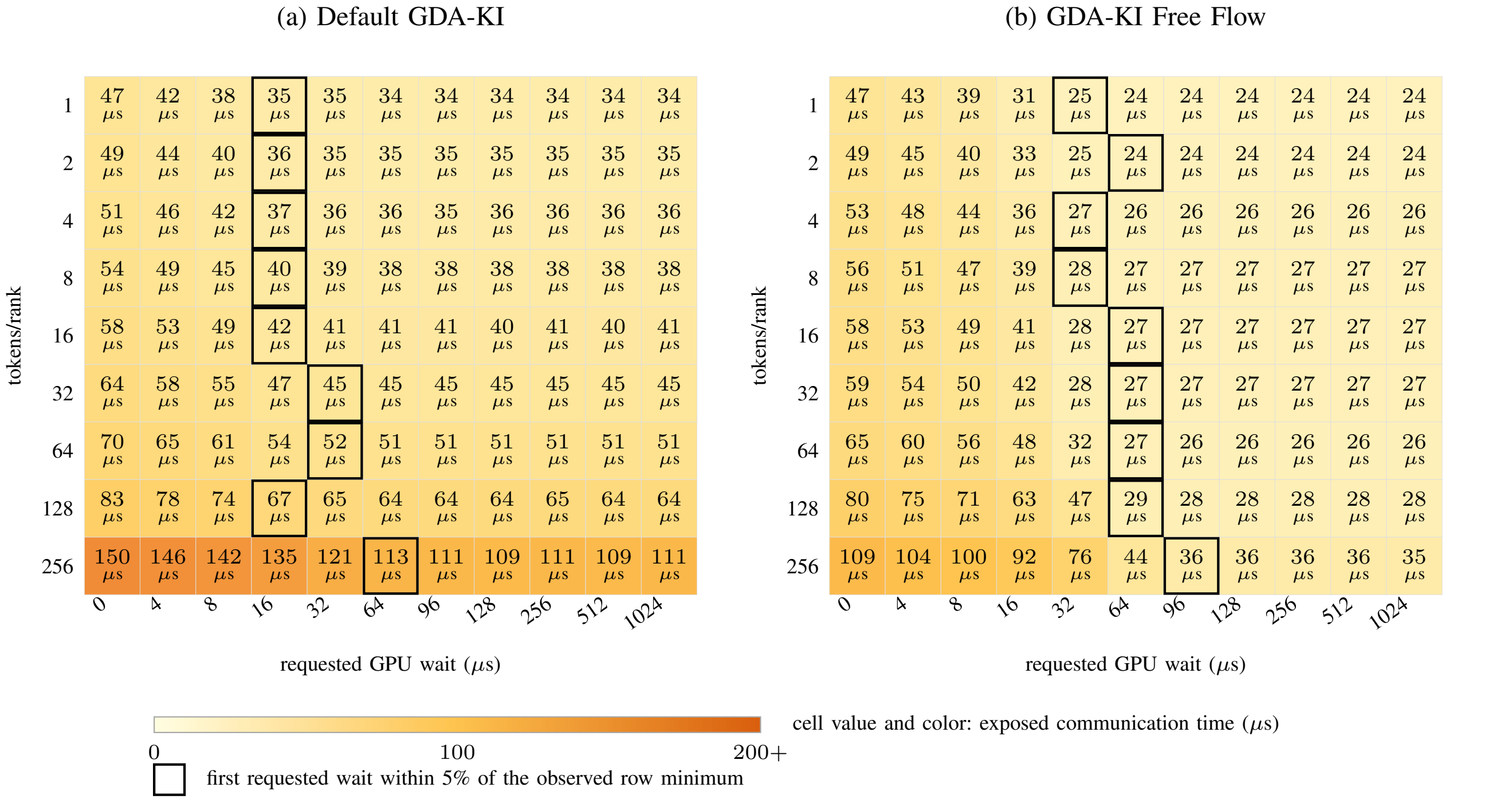

\paragraph{Dispatch-send reductions across the sweep}
Figures~\ref{fig:ncclep-dispatch-send} and~\ref{fig:ncclep-dispatch-topk4}
show a consistent reduction across the evaluated range.  The complete EP8,
top-$k=8$, cap-128 row illustrates the scaling.  At two tokens/rank, default
GDA-KI and GDA-KI Free Flow take 29.5 and 19.0~$\mu$s, respectively.  At
128 tokens those times become 57.5 and 18.9~$\mu$s; at 256 tokens they become
107.1 and 25.7~$\mu$s.

Across the completed factorial, GDA-KI Free Flow's mean paired dispatch-send
time change is negative in all 162 positions.  These are descriptive counts over heterogeneous shapes,
with five rounds per position, and do not by themselves establish a serving
gain.

\subsection{Exposed Communication Time under Controlled Waiting}
\label{sec:overlap-results}

\noindent\textbf{Every cell in Figure~\ref{fig:ncclep-overlap} is absolute
exposed communication time in $\mu$s.}

Figure~\ref{fig:ncclep-overlap} answers two questions: how long network
progress stays in flight after dispatch send releases the GPU, and how much
exposed time remains once that progress is hidden.  Along each row, exposed
time falls as the GPU wait grows, because receive waits less for in-flight
data, and then levels off once the wait covers the in-flight interval.  The
marked cell is therefore the covering window a backend needs: its in-flight
time is approximately that wait.  The level the row settles at, its floor,
is the send and receive work that remains on the GPU however long the
covering work is.

\begin{figure*}[!t]
\centering
\input{figures/serving_percentage}
\caption{Measured vLLM TPOT for nominal 128 input and 128 output tokens.
(a) The NCCL-EP baseline shows actual median TPOT in milliseconds; the
DeepEP baseline and optimized NCCL-EP rows show percentage changes from that
NCCL-EP baseline across the full concurrency
sweep.  Concurrency 1 is a stable non-DBO control.  \textbf{The hatched
concurrency-32/64 region is not transport evidence: variation in scheduled
decode batch sizes across ranks can fail the all-rank DBO vote on some
steps, and those non-DBO steps have no captured graph and run eagerly.  The
single graph per shape is intentional, so that this penalty exposes cells
that mix DBO and non-DBO steps (Section~\ref{sec:serving-method}).
Splitting roughly two tokens into one-token microbatches is also not an
economical DBO regime.}  (b) Absolute median TPOT for all
three series in the reliable-DBO region.  Both baseline series use default GDA-KI.
Every plotted value is the mean of three run medians, and every relative value uses the same three-run
NCCL-EP reference.
All configurations use default GDA-KI prefill.}
\label{fig:serving-concurrency}
\end{figure*}
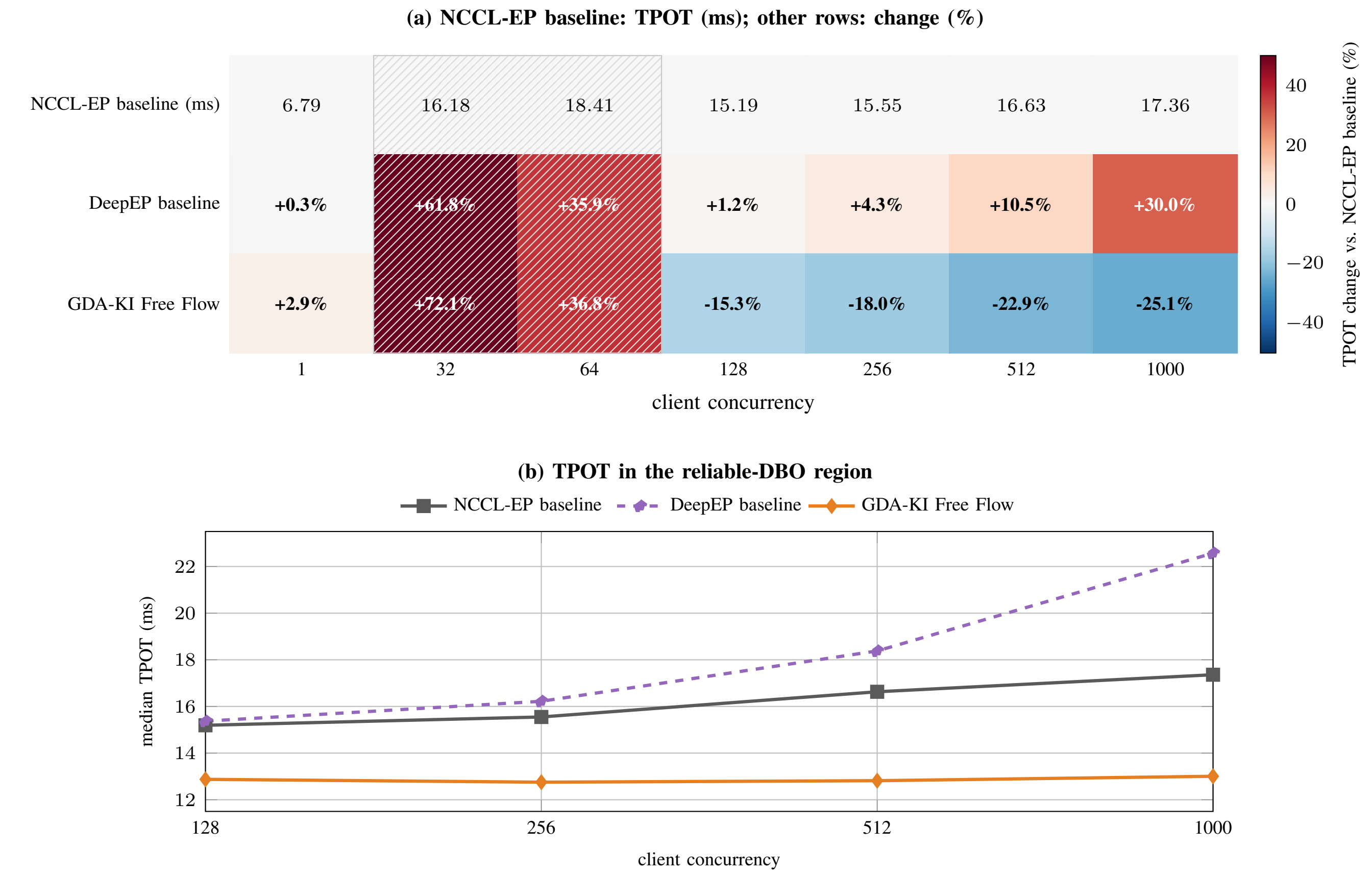

\paragraph{Offload lengthens the in-flight interval}
Default GDA-KI needs the shortest covering window: 16--32~$\mu$s up to
128 tokens/rank and 64~$\mu$s at 256.  Its GPU builds WQEs and rings
doorbells during send, so little work remains in flight after release.  GDA-KI
Free Flow moves that work out of send and into the in-flight interval, so it
needs a longer window.  It nevertheless keeps a short window
across the sweep: 32--64~$\mu$s up to 128 tokens/rank and 96~$\mu$s at 256.

\paragraph{Once hidden, offload leaves less exposed time}
With a long enough window, GDA-KI Free Flow settles at a lower floor than
default GDA-KI.  At 1--16 tokens/rank its floor is 24--27~$\mu$s, against
34--41~$\mu$s for default GDA-KI.  At 256 tokens/rank default GDA-KI still
exposes 108.8~$\mu$s, against 35.5~$\mu$s for Free Flow.

\paragraph{Implication}
The two NCCL-EP experiments answer complementary questions.  First, the
in-flight interval only has to fit within the covering work: once it does,
it is fully hidden, and a shorter interval brings no further gain.  Default
GDA-KI's short window is therefore an advantage only when the covering work
is shorter than Free Flow's 32--96~$\mu$s window.  Second, once the in-flight
interval is covered, exposed time is set by the send and receive work that
stays on the GPU, the floor in Figure~\ref{fig:ncclep-overlap}.  Its send
component is the dispatch-send time of Section~\ref{sec:dispatch-results},
which is why the path that releases the GPU earlier also reaches a lower
floor.  Section~\ref{sec:serving-results} tests whether DBO in vLLM decode
supplies enough covering work to hide the in-flight interval.

\subsection{vLLM TPOT}
\label{sec:serving-results}

Figure~\ref{fig:serving-concurrency}(a) reports the NCCL-EP default-GDA-KI
baseline in milliseconds and percentage changes for the DeepEP default-GDA-KI
baseline and the optimized NCCL-EP configuration, including the
low-concurrency controls.  Figure~\ref{fig:serving-concurrency}(b) reports
absolute TPOT for these three series in the concurrency-128--1000 region
where DBO operates reliably.

For each backend and concurrency we average three independent run-level
median TPOT values, then normalize to the same three-run NCCL-EP default-GDA-KI
mean.  Percentages are ratios of these means.

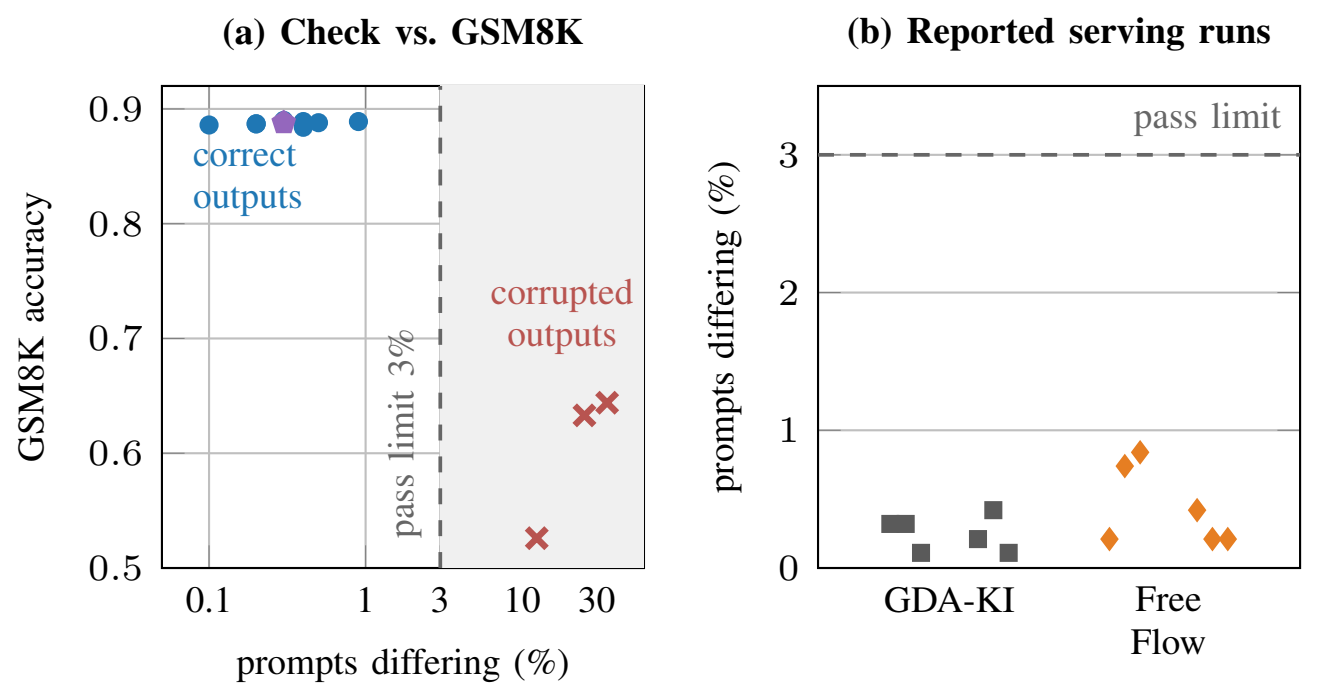
\begin{figure}[!t]
\centering
\input{figures/correctness}
\caption{Output correctness.  (a) GSM8K accuracy against the share of the
949 reference prompts whose output differs from DeepEP, for validation runs
of correct configurations (blue), DeepEP itself (purple), and configurations
with the dma-buf address-wrap bug (red).  (b) The same check for every
NCCL-EP serving run reported in this paper, three runs per backend for each
of the two sweeps.  The dashed line marks the 3\% pass limit.}
\label{fig:correctness}
\end{figure}

\paragraph{Correctness}
Every NCCL-EP serving run completes with zero failed requests and passes the
output check, with 0.11--0.84\% of the 949 reference prompts differing from
DeepEP (Figure~\ref{fig:correctness}(b)).  The check tracks task accuracy
(Figure~\ref{fig:correctness}(a)): on the GSM8K validation runs, correct
configurations scored 0.884--0.890, matching DeepEP (0.888) and a single GPU
(0.89), with at most 0.9\% differing, whereas configurations with the dma-buf
address-wrap bug scored 0.526--0.644 with 12.4--34.9\% differing.

\paragraph{Reliable-DBO region}
At concurrency 128--1000, GDA-KI Free Flow lowers median TPOT by
15.3--25.1\% relative to default GDA-KI.  The reduction grows with
concurrency.  At every one of these points, the slowest Free Flow run is
faster than the fastest default-GDA-KI run.

\paragraph{DeepEP baseline comparison}
At concurrency 128--1000, the DeepEP default-GDA-KI baseline is 1.2--30.0\%
slower than the NCCL-EP baseline, and the gap widens with concurrency.  Both
baselines are shown explicitly,
so the comparison separates the library baseline difference from the
gain of the optimized NCCL-EP path.

\paragraph{Low-concurrency controls}
\label{sec:low-concurrency}

At concurrency 1, vLLM never splits the decode batch.  CUDA-graph replay is
stable because the step is always unsplit, but there is no second microbatch
behind which to hide network progress.  Accordingly, GDA-KI Free Flow is
2.9\% slower than the GDA-KI baseline;
this cell does not test the DBO-overlap claim.

At concurrency 32--64, the decode tokens scheduled on each DP rank lie near
DBO's two-token threshold.  Variation in request placement and scheduled
batch sizes can make one step fail the all-rank vote, because some rank falls
below the threshold or because padding to the largest rank would leave the
final microbatch empty, selecting the non-DBO path, while another step passes
it and selects DBO.  Because we capture only one CUDA graph per padded shape,
in the split state, each non-DBO step falls back to eager decode.  This
large, backend-independent penalty is the intended signal of the
single-graph design (Section~\ref{sec:serving-method}): it marks these
cells as mixtures of DBO and non-DBO steps, which are unsuitable for
comparing network backends.  Capturing a second, unsplit graph would hide
the penalty but leave the mixture, so the measured TPOT could no longer be
attributed to DBO.

The regime is also not a useful target for DBO.  With roughly two tokens on
a rank, DBO creates two one-token microbatches on the same compute stream,
leaving almost no compute behind which to hide communication while still
paying the split, scheduling, and synchronization overhead.  We do not treat
such splits as an economical DBO configuration.  Concurrency 32--64 is
therefore excluded as transport evidence for two reasons: its steps mix
split and unsplit execution, and its near-threshold splits are too small to
benefit from overlap.

\paragraph{Boundary}
The result does not show that enabling DBO is always an absolute win.  It
requires enough per-rank work to amortize the split and expose a useful
covering window.  The supported result is narrower: at the concurrency points
where DBO splits reliably and economically, the optimized configuration
has lower median TPOT, consistent with earlier release improving overlap.
The serving experiment compares whole configurations; it does not isolate
dispatch-send duration, receive/combine cost, or final remote-arrival time.
Unchanged or slower network completion remains a conditional scheduling
argument, rather than a measured serving result.

\begin{figure}[!ht]
\centering
\input{figures/serving_output_sweep}
\caption{Measured vLLM output-length sweep at fixed concurrency 256 and
nominal input length 128.  Each value is the mean of three run-level median TPOTs,
with 1,024 measured requests per point.
Prefill is default GDA-KI in every configuration.  The additional DeepEP default-GDA-KI
baseline uses three runs per point.}
\label{fig:serving-output-length}
\end{figure}
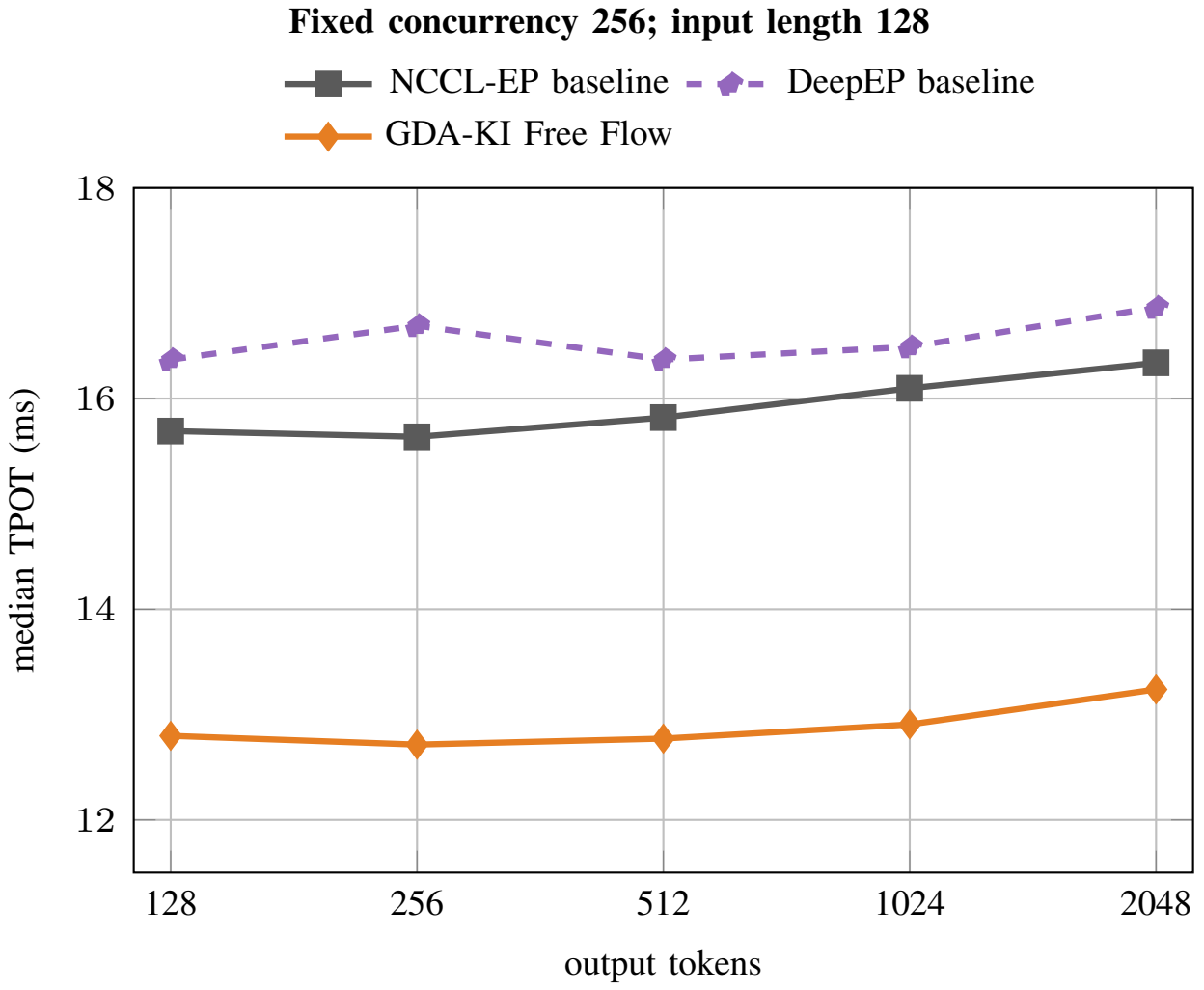

\paragraph{Output-length sensitivity}
Figure~\ref{fig:serving-output-length} holds client concurrency at 256 and
nominal input length at 128 tokens, then varies output length from 128 to
2048 tokens.  The NCCL-EP baseline ranges from 15.64 to 16.34~ms,
while GDA-KI Free Flow remains within 12.71--13.24~ms.  The optimized
configuration retains its median-TPOT advantage across the measured
output lengths.  The DeepEP baseline ranges from 16.37 to 16.87~ms, or 2.4--6.8\%
above the NCCL-EP baseline at the corresponding output lengths.

The feature therefore sustains its benefit for workloads with
reliable DBO and sufficient progress resources.

%% file: figures/ncclep_dispatch_send.tex
\centerline{\small\bfseries Top-$k=8$}
\vspace{2mm}
\begin{tabular}{@{}c@{\hspace{5mm}}c@{}}
\ncclepdispatchabsolute{figures/data/ncclep_dispatch_sweep_gdaki_topk8.csv}
  {(a) Default GDA-KI ($\mu$s)} &
\ncclepdispatchrelative{figures/data/ncclep_dispatch_sweep_free_flow_topk8.csv}
  {(b) GDA-KI Free Flow (\%)}
\end{tabular}

\vspace{2mm}
\input{figures/dispatch_send_legend}

%% file: figures/dispatch_send_legend.tex
\begin{tikzpicture}
  \shade[left color={rgb,255:red,247;green,251;blue,255},
         right color={rgb,255:red,8;green,48;blue,107}]
    (0,0) rectangle (3.2,0.18);
  \draw[black!35] (0,0) rectangle (3.2,0.18);
  \node[font=\scriptsize,anchor=north] at (0,-0.02) {$0$};
  \node[font=\scriptsize,anchor=north] at (1.6,-0.02) {$175$};
  \node[font=\scriptsize,anchor=north] at (3.2,-0.02) {$350~\mu$s};
  \node[font=\scriptsize,anchor=west] at (3.4,0.09) {panel (a)};
  \begin{scope}[xshift=7.0cm]
    \shade[left color={rgb,255:red,5;green,48;blue,97},
           right color={rgb,255:red,103;green,0;blue,31},
           middle color={rgb,255:red,247;green,247;blue,247}]
      (0,0) rectangle (4.0,0.18);
    \draw[black!35] (0,0) rectangle (4.0,0.18);
    \node[font=\scriptsize,anchor=north] at (0,-0.02) {$-100\%$};
    \node[font=\scriptsize,anchor=north] at (2.0,-0.02) {$0\%$};
    \node[font=\scriptsize,anchor=north] at (4.0,-0.02) {$+100\%$};
    \node[font=\scriptsize,anchor=west] at (4.2,0.09)
      {(b); blue is better};
  \end{scope}
\end{tikzpicture}

%% file: figures/ncclep_full.tex
\begin{tabular}{@{}c@{\hspace{5mm}}c@{}}
\waitexposurematrix{figures/data/ncclep_wait_gdaki_matrix.csv}
  {(a) Default GDA-KI}{3/0,3/1,3/2,3/3,3/4,4/5,4/6,3/7,5/8} &
\waitexposurematrix{figures/data/ncclep_wait_free_flow_matrix.csv}
  {(b) GDA-KI Free Flow}{4/0,5/1,4/2,4/3,5/4,5/5,5/6,5/7,6/8}
\end{tabular}

\vspace{1mm}
\begin{tikzpicture}
  \shade[left color={rgb,255:red,255;green,255;blue,229},
         right color={rgb,255:red,217;green,95;blue,14},
         middle color={rgb,255:red,254;green,196;blue,79}]
    (0,0) rectangle (7.0,0.18);
  \draw[black!35] (0,0) rectangle (7.0,0.18);
  \node[font=\scriptsize,anchor=north] at (0,-0.02) {$0$};
  \node[font=\scriptsize,anchor=north] at (3.5,-0.02) {$100$};
  \node[font=\scriptsize,anchor=north] at (7.0,-0.02) {$200+$};
  \node[font=\scriptsize,anchor=west] at (7.25,0.09)
    {cell value and color: exposed communication time ($\mu$s)};
  \draw[black,line width=0.8pt] (0,-0.72) rectangle (0.35,-0.37);
  \node[font=\scriptsize,anchor=west] at (0.48,-0.545)
    {first requested wait within 5\% of the observed row minimum};
\end{tikzpicture}

%% file: figures/serving_percentage.tex
\centerline{\footnotesize\bfseries
(a) NCCL-EP baseline: TPOT (ms); other rows: change (\%)}
\vspace{0.5mm}
\servingmatrix{figures/data/serving_full_matrix.csv}

\vspace{3mm}
\centerline{\footnotesize\bfseries
(b) TPOT in the reliable-DBO region}
\vspace{0.5mm}
\input{figures/serving_tpot}

%% file: figures/correctness.tex
\begin{tikzpicture}
\begin{axis}[
  name=validation,
  title={(a) Check vs.\ GSM8K},
  title style={font=\scriptsize\bfseries,yshift=-1mm},
  width=0.36\columnwidth,
  height=0.36\columnwidth,
  scale only axis,
  xmode=log,
  xmin=0.05,xmax=60,ymin=0.5,ymax=0.92,
  xtick={0.1,1,3,10,30},
  xticklabels={0.1,1,3,10,30},
  ytick={0.5,0.6,0.7,0.8,0.9},
  xlabel={prompts differing (\%)},
  ylabel={GSM8K accuracy},
  label style={font=\scriptsize},
  tick label style={font=\scriptsize},
  grid=major,
  clip=false,
]
\fill[black!6] (axis cs:3,0.5) rectangle (axis cs:60,0.92);
\draw[black!60,dashed,line width=0.7pt] (axis cs:3,0.5) -- (axis cs:3,0.92);
\node[font=\scriptsize,text=black!60,anchor=south west,rotate=90]
  at (axis cs:3,0.51) {pass limit 3\%};
\addplot[only marks,mark=*,color=correctcolor,mark size=1.6pt]
  table[col sep=comma,x=dev_pct,y=gsm8k]
  {figures/data/correctness_validation_correct.csv};
\addplot[only marks,mark=pentagon*,color=deepepcolor,mark size=2.2pt]
  table[col sep=comma,x=dev_pct,y=gsm8k]
  {figures/data/correctness_validation_deepep.csv};
\addplot[only marks,mark=x,color=gpudraw,mark size=2.6pt,very thick]
  table[col sep=comma,x=dev_pct,y=gsm8k]
  {figures/data/correctness_validation_corrupted.csv};
\node[font=\scriptsize,text=gpudraw,anchor=south,align=center]
  at (axis cs:18,0.67) {corrupted\\outputs};
\node[font=\scriptsize,text=correctcolor,anchor=south west,align=left]
  at (axis cs:0.06,0.79) {correct\\outputs};
\end{axis}

\begin{axis}[
  at={(validation.east)},anchor=west,xshift=0.13\columnwidth,
  title={(b) Reported serving runs},
  title style={font=\scriptsize\bfseries,yshift=-1mm},
  width=0.36\columnwidth,
  height=0.36\columnwidth,
  scale only axis,
  xmin=-0.6,xmax=1.6,ymin=0,ymax=3.5,
  xtick={0,1},
  xticklabels={GDA-KI,{Free\\Flow}},
  xticklabel style={font=\scriptsize,align=center,anchor=north},
  xtick style={draw=none},
  ytick={0,1,2,3},
  ylabel={prompts differing (\%)},
  label style={font=\scriptsize},
  tick label style={font=\scriptsize},
  ymajorgrids=true,
  clip=false,
]
\draw[black!60,dashed,line width=0.7pt] (axis cs:-0.6,3) -- (axis cs:1.6,3);
\node[font=\scriptsize,text=black!60,anchor=south east]
  at (axis cs:1.6,3) {pass limit};
\addplot[only marks,mark=square*,color=gdakicolor,mark size=1.5pt]
  table[col sep=comma,x=x,y=dev_pct,
        restrict expr to domain={\thisrow{x}}{-0.5:0.5}]
  {figures/data/correctness_runs.csv};
\addplot[only marks,mark=diamond*,color=freeflowcolor,mark size=2.0pt]
  table[col sep=comma,x=x,y=dev_pct,
        restrict expr to domain={\thisrow{x}}{0.5:1.5}]
  {figures/data/correctness_runs.csv};
\end{axis}
\end{tikzpicture}

%% file: figures/serving_output_sweep.tex
\centerline{\scriptsize\bfseries
Fixed concurrency 256; input length 128}
\vspace{0.5mm}
\begin{tikzpicture}
\begin{axis}[
  width=0.82\columnwidth,
  height=0.53\columnwidth,
  scale only axis,
  xmin=-0.15,xmax=4.15,ymin=11.5,ymax=18,
  xtick={0,1,2,3,4},
  xticklabels={128,256,512,1024,2048},
  xlabel={output tokens},
  ylabel={median TPOT (ms)},
  label style={font=\scriptsize},
  tick label style={font=\scriptsize},
  grid=major,
  legend columns=2,
  legend style={font=\scriptsize,draw=none,at={(0.5,1.02)},anchor=south},
]
\addplot[color=gdakicolor,mark=square*,very thick]
  table[col sep=comma,x=x,y=gdaki_tpot_ms]
  {figures/data/serving_output_sweep.csv};
\addlegendentry{NCCL-EP baseline}

\addplot[color=deepepcolor,mark=pentagon*,very thick,dashed]
  table[col sep=comma,x=x,y=deepep_gdaki_tpot_ms]
  {figures/data/serving_output_sweep.csv};
\addlegendentry{DeepEP baseline}

\addplot[color=freeflowcolor,mark=diamond*,very thick]
  table[col sep=comma,x=x,y=free_flow_tpot_ms]
  {figures/data/serving_output_sweep.csv};
\addlegendentry{GDA-KI Free Flow}
\end{axis}
\end{tikzpicture}

%% file: sections/08_discussion.tex
\section{Discussion}
\label{sec:discussion}

\subsection{Optimize Exposed Communication Time}

The experiments support a scheduling principle:
\begin{quote}
\emph{Minimize exposed communication time by releasing the GPU early, then
assign remaining progress to an engine that can run beneath useful work.}
\end{quote}
This objective is more precise than minimizing one isolated transport call or
the sequential dispatch/combine total.  GDA-KI Free Flow leaves WQE
construction on the GPU but moves ready-prefix serialization and doorbells to
a host progress thread.  The controlled wait
sweep shows the progress opportunity after release, while the serving
experiment measures the TPOT of the complete configurations with a common
default-GDA-KI prefill backend.

The principle allows an initially counterintuitive trade.  Suppose backend
$b$ releases the GPU $\Delta$ earlier but has the same or slightly longer time
until the final remote arrival.  By (\ref{eq:release-bound}), work from the
other microbatch starts up to $\Delta$ earlier, and as long as the final
remote arrival falls within that covering work, receive does not wait on the
network and the layer finishes earlier.  In practice the covering work is
long enough: the Free Flow in-flight windows in
Section~\ref{sec:overlap-results} are 32--96~$\mu$s, and in vLLM decode Free
Flow lowers TPOT at every reliable-DBO point.  Under DBO, dispatch-send time, not network
completion time, is therefore the quantity to minimize.

\paragraph{The optimization objective determines the preferred path.}
These results are not a universal ranking of GIN backends.  Under DBO, the
objective is to minimize exposed communication time by releasing the GPU
early enough for another microbatch to run, while network-side progress
continues beneath that work.  Without DBO or another source of covering
work, the relevant objective becomes the end-to-end latency from dispatch
send through in-flight progress to dispatch receive.  We do not evaluate
that regime, and the preferred backend there may differ.  The evaluated
prototype therefore uses GDA-KI Free Flow for host-assisted progress.
Selection depends on workload, available host progress resources, and which
communication intervals the application can hide.

\subsection{Limitations}

\textbf{GPU wait as covering work.}
In a real workload, the other microbatch launches expert GEMMs between
dispatch send and receive.  In time, the requested wait stands in for that
covering work: a GEMM that lasts as long as the wait gives the network the
same progress window, plus some launch and scheduling overhead.

\textbf{Serving controls.}
Each backend and point has three runs.  The serving experiment swaps
complete backends, so a TPOT change reflects the whole backend, including
its receive and combine paths, not dispatch send alone; the native
experiments identify earlier GPU release with hidden in-flight progress as
the mechanism behind it.  DeepEP appears only as a default-GDA-KI serving
baseline.

\textbf{Topology and scale.}
The native experiments cover 4, 8, and 16 GPUs over one, two, and four
nodes, and the serving experiments use two four-GPU decode nodes serving
Qwen3-30B-A3B with 128 input tokens, up to 2,048 output tokens, and client
concurrency up to 1,000.  Production deployments differ along every one of
these axes: longer inputs and outputs, decode groups of 20 or more nodes,
higher concurrency, and larger models.  Each changes the share of TPOT that
dispatch communication occupies, and therefore how much earlier GPU release
matters.  Larger decode groups spread tokens across more remote peers and
may make network progress a larger share of each step; larger models and
longer sequences add expert and attention compute that may make it a
smaller share.  We do not measure which effect dominates, so the gains
reported here may grow or shrink in production.  Intranode-only NVLink is
a deliberate mixed path and does not model a fully connected NVL72
deployment.

\textbf{CPU progress capacity.}
GDA-KI Free Flow depends on a host thread that continuously polls for ready
WQEs.  Its gains therefore assume sufficient host CPU capacity for
timely progress.  In CPU-intensive serving pipelines---including agentic
workloads with substantial host-side orchestration, tokenization, retrieval,
or tool-processing overhead---contention for that CPU capacity may delay
network progress, reduce the observed benefit, or limit scaling.  We did not
evaluate CPU-saturated conditions or affinity sensitivity, so the results
should not be extrapolated to those regimes.

\textbf{PCIe polling contention.}
On systems where the host reads GPU-resident readiness state across PCIe,
Free Flow polling adds PCIe transactions.  When PCIe is already busy, those
transactions add pressure and may take longer to complete, further delaying
network progress.  We did not evaluate PCIe contention or sensitivity to
ready-state memory placement, so the results should not be extrapolated to
systems with a saturated host--device path.

\textbf{The other three communication phases.}
This paper optimizes only dispatch send.  Exposed communication time also
includes dispatch receive, combine send, and combine receive, and each can
be reduced in its own way: combine send can apply the same early-release
principle, while the two receive phases determine how much GPU work remains
once data arrives.  Optimizing all four toward minimal exposed
communication time is future work.

\textbf{Alternative schedules.}
All serving results use single-stream DBO: both microbatches and NCCL-EP
send and receive share the caller's compute stream.  We did not compare
against two other schedules: running communication, or each microbatch, on
its own CUDA stream, and serving without DBO.  Either could be faster than
single-stream DBO for some workloads.  If so, the earlier release we
optimize for would matter less, so both comparisons need future
experiments.

%% file: sections/09_related_work.tex
\section{Related Work}
\label{sec:related}

\textbf{GPU-initiated networking.}
GPUrdma and GPUDirect Async remove the CPU from the communication critical
path \cite{gpurdma,agostini2018gpudirectasync}.  GPU-initiated OpenSHMEM and
NVSHMEM provide in-kernel one-sided communication and collectives
\cite{hamidouche2020gpuopenshmem,nvshmem}, DOCA GPUNetIO exposes NIC queues
and doorbells to GPU kernels \cite{docaGpuNetIO}, and NCCL GIN integrates
device-initiated operations with NCCL communicators
\cite{hamidouche2025gin,ncclDeviceApi}.  Groves et al.\ compare
device-initiated and host-initiated NVSHMEM communication
\cite{groves2020tradeoffs}.  Baydamirli et al.\ dissect GPU-initiated
communication at the GPU--NIC boundary across NVSHMEM, NCCL GIN, DeepEP, and
other libraries, finding that libraries add up to 4.6~$\mu$s of issue time
through queue management, memory ordering, and completion scope, and that a
tuned CPU proxy matches the GPU path at idle at the cost of a dedicated core
\cite{baydamirli2026dissecting}.  These studies characterize the submission
path itself; we ask how its placement changes exposed communication time
under DBO.

\textbf{Offloading communication progress.}
BluesMPI offloads the progress of non-blocking MPI Alltoall to BlueField DPU
cores to free the host for computation \cite{bayatpour2021bluesmpi}.
UCCL-EP keeps GPU-initiated token routing but delegates RDMA issue to CPU
proxy threads for portability across GPUs and NICs \cite{mao2026uep}.
GDA-KI Free Flow follows the same principle of moving progress off the
critical resource, applied to the GPU SM by moving send progression to a
host thread.

\textbf{MoE communication and overlap.}
Sparse MoE models route each token to experts sharded across devices,
creating data-dependent all-to-all traffic
\cite{shazeer2017moe,lepikhin2020gshard,fedus2022switch}.  FasterMoE
replicates popular experts and overlaps fine-grained all-to-all with expert
computation \cite{fastermoe}, and Tutel adapts parallelism and pipelining at
runtime and optimizes all-to-all \cite{tutel}.  A line of work hides MoE
communication behind computation: Lancet overlaps all-to-all with the whole
training graph \cite{jiang2024lancet}; Comet and MegaScale-MoE fuse
tile-level communication into MoE compute kernels
\cite{zhang2025comet,jin2026megascalemoe}, and TileLink generalizes such
fusion into a compiler for overlapped compute--communication kernels
\cite{zheng2025tilelink}; and FlashMoE runs dispatch, expert computation, and
combine in a single persistent kernel \cite{aimuyo2025flashmoe}.  DeepSeek-V3
introduces specialized cross-node dispatch and combine kernels that overlap
with computation \cite{deepseekV3}, and DeepEP releases such kernels as a
library whose low-latency mode adds receive hooks for overlap without
occupying SMs \cite{deepEP}.  NCCL-EP provides the expert-parallel interface
over NCCL used here \cite{ncclEp}.  These systems change \emph{what}
overlaps with communication; we keep the overlap schedule fixed and reduce
the GPU time that dispatch send itself consumes, which is complementary.

\textbf{Microbatch overlap.}
vLLM manages the KV cache with PagedAttention and uses iteration-level
(continuous) batching \cite{vllm}, and disaggregated systems separate prefill
and decode \cite{distserve,mooncake}.  In training, DeepSeek-V3's DualPipe
pairs a forward and a backward chunk, each split into attention, all-to-all
dispatch, MLP, and all-to-all combine, so that one chunk's all-to-all runs
beneath the other's computation \cite{deepseekV3}.  In serving, DeepSeek's
dual-batch overlap \cite{deepseek2025inference}, MegaScale-Infer's ping-pong
pipeline between attention and expert nodes \cite{zhu2025megascaleinfer},
and NanoFlow's nano-batching \cite{zhu2025nanoflow} similarly split a batch
so that one part's communication or memory-bound work runs beneath another
part's computation.
Our results show that, within such a schedule, the time communication holds
the GPU before releasing it determines how much of the overlap is realized.

%% file: sections/10_conclusion.tex
\section{Conclusion}
\label{sec:conclusion}

MoE communication should be optimized at the boundary the scheduler sees:
how long dispatch send holds the GPU.  GDA-KI Free Flow moves send progression
to a host thread and releases the GPU earlier than default GDA-KI across the
entire measured sweep.
Offload lengthens the in-flight interval, but under DBO the other microbatch's
work covers it, and what remains exposed is the shorter GPU-side work.  In
vLLM decode, this lowers median TPOT by 15--25\% relative to NCCL-EP default
GDA-KI and by 16--42\% relative to DeepEP v1 wherever DBO splits reliably.
The result demonstrates how host-assisted progress can minimize exposed
communication time under DBO.

%% file: sections/11_appendix.tex
\appendices
\section{Top-$k=4$ Dispatch-Send Results}
\label{app:dispatch-topk4}

\begin{figure}[!ht]
\renewcommand{\thefigure}{5b}
\centering
\input{figures/ncclep_dispatch_topk4}
\caption{Five-round NCCL-EP dispatch-send sweep at top-$k=4$.
The two-panel layout matches Figure~\ref{fig:ncclep-dispatch-send}:
(a) default-GDA-KI time in $\mu$s; (b) paired changes for GDA-KI Free Flow,
with \% signs in every relative cell.  All 162 plotted cells
have five matched rounds.  Colors saturate at $\pm100\%$ for relative
panels; printed values retain larger changes.  NVLink is disabled.}
\label{fig:ncclep-dispatch-topk4}
\end{figure}
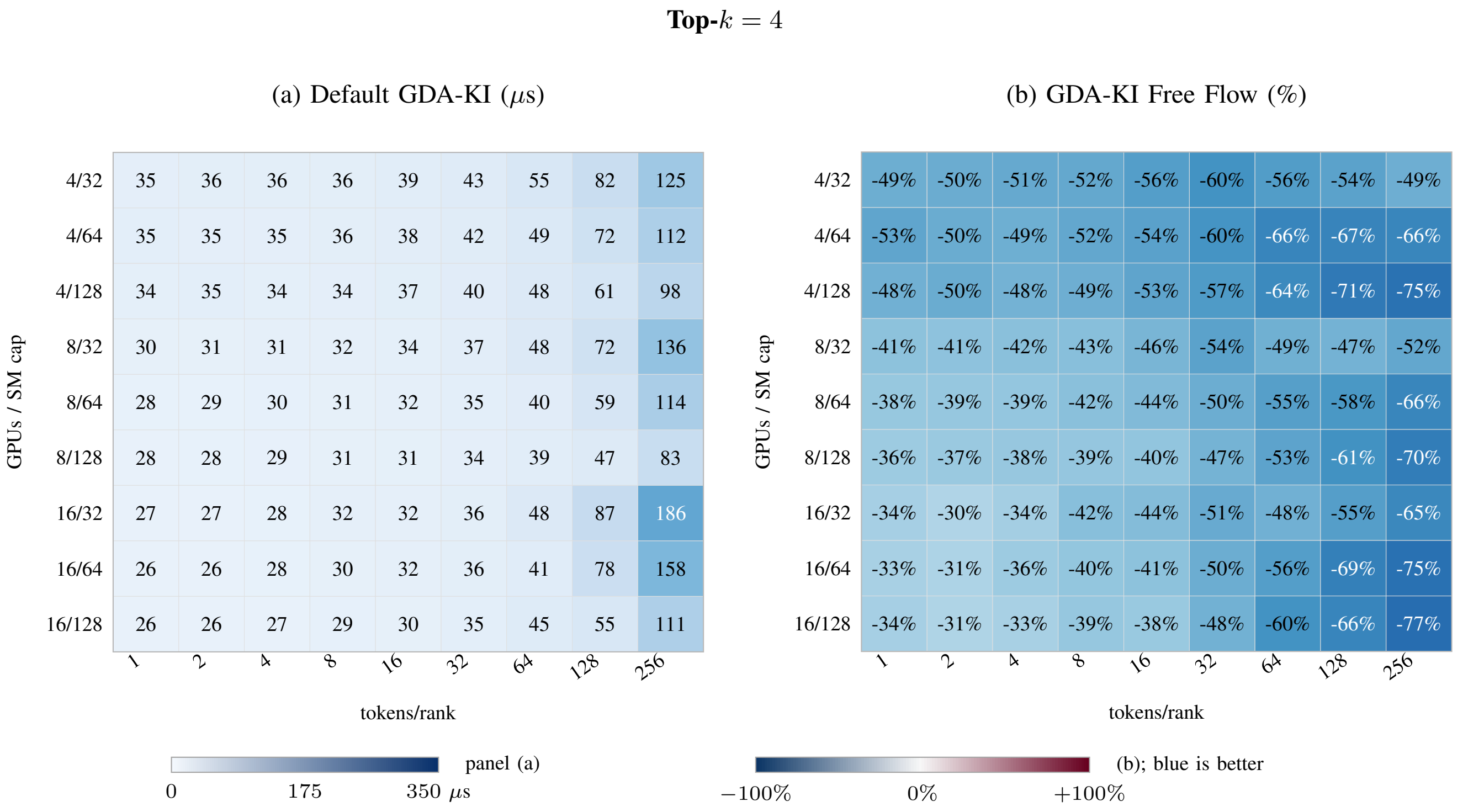

Figure~\ref{fig:ncclep-dispatch-topk4} completes the dispatch factorial with
its top-$k=4$ slice.  It uses the same five-round estimator, axes, colors,
and cell units as the top-$k=8$ result in Figure~\ref{fig:ncclep-dispatch-send}.

%% file: figures/ncclep_dispatch_topk4.tex
\centerline{\small\bfseries Top-$k=4$}
\vspace{2mm}
\begin{tabular}{@{}c@{\hspace{5mm}}c@{}}
\ncclepdispatchabsolute{figures/data/ncclep_dispatch_sweep_gdaki_topk4.csv}
  {(a) Default GDA-KI ($\mu$s)} &
\ncclepdispatchrelative{figures/data/ncclep_dispatch_sweep_free_flow_topk4.csv}
  {(b) GDA-KI Free Flow (\%)}
\end{tabular}

\vspace{2mm}

\input{figures/dispatch_send_legend}